\documentclass[superscriptaddress,reprint,amsmath,amssymb,aps]{revtex4-2}
\usepackage[utf8]{inputenc}
\usepackage{graphicx}
\usepackage{amsmath}
\usepackage{enumerate}
\usepackage[T1]{fontenc}
\usepackage{tipa}
\usepackage{xcolor}
\newcommand{\avg}[1]{\left \langle #1  \right\rangle}
\usepackage[version=4]{mhchem}

\usepackage{ulem}
\usepackage{xcolor}

\newcommand{\agregar}[1]{{\color{blue}#1}}

\begin{document}

\preprint{APS/123-QED}

\title{A model for phonetic changes driven by social interactions}

\author{A. Chacoma}
\email{achacoma@famaf.unc.edu.ar}
\affiliation{Instituto de F\'isica Enrique Gaviola (IFEG-CONICET).}

\author{N. Almeira}
\affiliation{Instituto de F\'isica Enrique Gaviola (IFEG-CONICET).}
\affiliation{Facultad de Matem\'atica, Astronom\'ia, 
F\'isica y Computaci\'on, Universidad Nacional de C\'ordoba.}

\author{J.I. Perotti}
\affiliation{Instituto de F\'isica Enrique Gaviola (IFEG-CONICET).}

\author{O.V. Billoni}
\affiliation{Instituto de F\'isica Enrique Gaviola (IFEG-CONICET).}
\affiliation{Facultad de Matem\'atica, Astronom\'ia, 
F\'isica y Computaci\'on, Universidad Nacional de C\'ordoba.}

\begin{abstract}
We propose a stochastic model to study phonetic changes as 
an evolutionary process driven 
by social interactions between 
two groups of individuals with 
different phonological systems.
Particularly, we focus on the changes in 
the place of articulation, inspired by 
the drift /\textphi/$\rightarrow$/h/ observed in 
some words of Latin root in the Castilian language.
In the model, each agent is 
characterized by a variable of 
three states, 
representing the place of articulation 
used during speech production. 
In this frame, we propose stochastic rules of interactions
among agents which lead to phonetic imitation 
and consequently to changes in the articulation place.
Based on this, we mathematically 
formalize the model as a problem 
of population dynamics, 
derive the equations of evolution in 
the mean field approximation, and study 
the emergence of three non--trivial global states,
which can be linked to the pattern of 
phonetic changes observed in 
the language of Castile and in 
other Romance languages.
\end{abstract}


\maketitle

\section{Introduction}

Oral communication, as the process of transmitting
concepts and ideas from one individual to another
by word of mouth, has been a main feature of human kind
since the first primitive societies. 
Historically, research in this field
has been faced by anthropologists and linguistics. 
However, in the last years 
the interest for the development
of new technologies related to 
automatic speech recognition,
specially in artificial intelligent systems
\cite{battaglia2018relational}, has made
it an active multidisciplinary area
of research
\cite{walker2018speech,fitch2018biology,
daly2018physiology,sun2019modeling}.

Particularly, in the field 
of the physics of complex systems,
the topic has been faced 
from the point of view of 
competition and evolution. 
In ref. \cite{castello2008modelling}, for instance,
by performing agent based model simulations,
Castell\'o et al. analyze the competition between 
two socially equivalent languages, and study 
the dynamics in structured populations in 
the frame of complex network theory.
Similarly, in ref. \cite{stauffer2005microscopic}, 
Stauffer et al. focus on the concepts of evolution,
analyzing the rise and fall of languages
using both macroscopic differential equations 
and microscopic Monte Carlo simulations.
In ref. \cite{baronchelli2006sharp}
likewise, Baronchelli et al. focus on
the analysis of the emergence of grammatical constructions,
reporting an order/disorder transition where the system 
goes through a sharp symmetry breaking process to reach a 
shared set of conventions.
Moreover, the co-evolution of symbols and meanings has been studied
through elementary language games by Puglisi et al. 
in ref.~\cite{puglisi2008PNAS}
showing the emergency of a hierarchical category structure.

In the frame of linguistic theories,  
on the other hand, speech production might be thought as 
the combination of several cognitive processes:
the selection of the proper words
to express an idea, the suitable 
choice of a grammatical form, 
and the production of sounds via 
the motor system and the vocal apparatus
\cite{liebermann2019speech}.
In this work we focus on the latter, 
therefore, in the following we describe the
most relevant concepts regarding the 
production of sounds.

Formally, phonemes are the minimal units 
of either vocalic or consonant sounds 
needed to produce words.
In this regard, the set of phonemes 
which encompasses all the sounds needed to produce 
every word in a given language, define a 
phonological system (PS).
It is particularly important to emphasize that
phonemes are not sounds but 
formal abstractions of speech sounds.
Any phoneme in a PS might be a representation 
for a family of sounds, technically 
called phones which are recognized by speakers 
and linked to a specific sound during 
oral communications 
\cite{Smith2000phonetics}.
Physiologically,
the process by which the vocal 
apparatus produces sounds
is called phonation \cite{gordon2001phonation}.
Through this mechanism,
humans are able to
produce a wide range of sounds,
usually divided into two groups: 
vowels and consonants \cite{ladefoged2012vowels}. 
Let us focus on 
consonants production.
In this case,
the phonetic apparatus
uses a combination of tongue,
lips, teeth and the soft palate,
in order to shape the different air obstructions
needed to produce the sounds.
The point inside the vocal cavity where
the obstruction occurs is called 
articulation place (AP), 
and the manner in 
which the obstruction is shaped is
called articulation mode (AM).
This two dimensions, AP and AM,
are commonly used to 
categorize the main features of 
the consonants in a particular PS.

Since languages are in continuous
evolution, it is well known 
that under certain conditions,
this process might lead to changes into the PS 
\cite{harrington2019phonetic}.
Particularly, 
in this work we are interested 
in studying phonetic 
changes in consonants
related to variations in the AP.
In this regard,
it has been observed that
these changes are enhanced 
when two or more groups of people 
with different languages are forced to 
socially interact 
\cite{lieberman2007quantifying, 
gregory2018language, blevins2004evolutionary,
beddor2007applying, beddor2009coarticulatory}.
For instance, when
a group invades other group's territory,
or when two groups establish economic relations
(trade, exchange of services, etc).
The linguistic, phonetic and grammatical
mutual influences produced by 
the interaction among the groups,
define a linguistic stratum (LS) 
\cite{hockett1961linguistic, trivino2013towards}, 
where the persistent social interaction over time,
in a process of oral communication,
guides the evolution to a common PS
and to a common new language.

A notable example of 
phonetic change in the AP due to a LS,
and the main inspiration of this work,
is the case of the glottalization
of the bilabial fricative 
phoneme /\textphi/ towards 
the glottal fricative /h/
(hereafter referred to as the change /\textphi/$\rightarrow$/h/),
and its subsequent disonorisation 
in some Latin root words of Castilian language 
(see table \ref{ta:words}).
It is thought that
the social process which led to the LS 
in this particular case,
was related to the social interactions among 
the prehispanic tribes (Iberics, Asturians, 
and mainly Vascons)
and the Romans, 
which were forced to socially 
interact in the Iberian peninsula---during the period of Rome's domain---from the second to the ninth cenrury, AD
\cite{lloyd1987latin,
ball2012celtic,bengtson2011basque}.
In this particular LS there were 
groups of people with a PS 
(based on Celtic language),
socially interacting with 
other group of people with a total 
different PS (based on Latin language).

In this context, it is thought that
the change /\textphi/$\rightarrow$/h/  
is related to prehispanic tribes speakers
performing changes in the AP during 
fricative production, trying to improve 
their communication skills with Latin speakers
\cite{flege2003interaction}.
Notably, these changes are not observed
in other Romance languages 
on the Iberian peninsula,
as the case of Portuguese
or Catalan, which emerge from a similar
LS than Castilian.
This fact led to researchers to theorize
about the properties of this particular LS, and
additionally to propose alternative theories
\cite{foulkes1997historical,salvador1983hipotesis}.
Until now, it seems there is not 
a total consensus regarding
the causes which led to these different 
evolution\agregar{s} of the Romance languages,
therefore the case is currently considered 
by linguistics as an open problem.

Motivated on the historical observations regarding the 
changes /\textphi/$\rightarrow$/h/, the aim of this work is
to propose a model of language competition 
\cite{abrams2003linguistics,nowak2002computational,baronchelli2007role,baronchelli2008depth,baxter_blythe_croft_mckane_2009} which captures the internal dynamics of a
LS. The model consider two social groups having different PS, 
where the changes in the articulation places are guided by social interactions 
based on rules of phonetic imitation.

We face the problem
in the frame of population dynamics,
where we study the evolution
of the changes in both groups,
and the emergence of general states
of pronunciation in the LS.

This paper is divided into three
main sections: 
in section \ref{sec:model}
we mathematically formalize the
model, define the main variables,
propose the rules of the interactions
and describe the dynamics; 
in section \ref{sec:eq_ev}
we derive from first principles 
the equations of evolution;
and, lastly, in section \ref{sec:estabilidad}
we analyze the emergence of global
states by performing an analysis
of both the evolution equations
and agent based model simulations.
We found that by tuning
the parameters related to 
the social interactions in the LS, 
our model shows the emergence
of three general global states
which capture qualitatively
the observations of
the emergent Romance languages 
of the Iberian peninsula, reinforcing
from our mathematical approach,
the stratum-based theories present 
in the literature.

\section{The model}
\label{sec:model}

We aim to model a process 
of phonetic imitation which leads
to changes in the AP during consonant
production.
Accordingly, we have made the following 
simplifications: we will
(i) limit our analysis to the 
changes in the AP, neglecting any 
change in the AM;
(ii) propose there are only 
three possible AP in the vocal cavity,
a \textit{front} place (bilabial, labio-dental),
a \textit{middle} place (dental, alveolar 
and post-alveolar) and a \textit{back} 
place (palatal, velar, uvular and glottal);
(iii) suppose there are two PS
in the LS, one which 
favors front and middle 
production, and other 
which favor middle and back; and
(iv) study the evolution of the changes in
the pronunciation of a single word.

In this frame, we define the main
elements of the model as follows: 

\begin{itemize}
    \item $A,B$ are two groups of agents in a stratum $LS$;
    
    \item $N_A$, $N_B$ are the number of agents in 
    $A,B$ and $N=N_A+N_B$ the total
    number of agents in $LS$;
    
    \item $S$ is the state of an agent in the $LS$ at time $t$,
    where $S \in \{1,2,3\}$ represents 
    the AP of agent $i$, such that:
    $1=front$, $2=middle$, $3=back$;
    
    \item $PS^A, PS^B$ are the phonological systems of $A,B$,
    indicating that $A$ has $S=1,2$ as preferential states
    (front-- middle); and conversely, $B$ has $S=2,3$ 
    (middle-- back).

\end{itemize}


\begin{table}
\centering
\begin{tabular}{|c|c|c|c| }
\hline
Latin & Castilian & Portuguese & English Translation \\
\hline
\textit{facere} & hacer  & fazer  & to do/ to make\\
\textit{femina} & hembra & fêmea & female \\
\textit{ferru}  & hierro & ferro & iron \\
\textit{filiu}  & hijo   & filho & son \\
\textit{folia}  & hoja   & folha & leaf \\
\textit{fumu}   & humo   & fumaça & smoke \\
\hline
\end{tabular}
\caption{
Some examples of Latin root words which  
show the change /\textphi/$\rightarrow$/h/ in Castilian (column 1).
In column 2, we show the translation into Portuguese
in order to show the change in this case did not happen. 
In column 3, we show the translation into English as a reference.
In order to clarify, note that in column 1, the phonetic transcription
of letter \textit{f} is the bilabial phoneme /\textphi/,
and in column 2, the phonetic transcription of letter \textit{h} is
the glottal phoneme /h/.}
\label{ta:words}
\end{table}


In the evolutionary process,
at time $t$, we randomly take
from the set $A\cup B$
an active agent and a reference agent.
The state of the former will change according
to the state of the latter,
guided by the following imitation rules
which we summarize by using 
the usual chemical reactions notation,

\begin{subequations} \label{eq:reactions}
\begin{align} 
\ce{A_2 + B_3 &->[q] A_3 + B_3} \label{rule1}\\
\ce{B_2 + A_1 &->[q] B_1 + A_1} \label{rule2}\\
\ce{A_3 + A_2 &->[p] 2A_2} \label{rule3} \\
\ce{B_1 + B_2 &->[p] 2B_2} \label{rule4}\\
\ce{A_1 &<=>[r] A_2} \label{rule5}\\
\ce{B_2 &<=>[r] B_3} \label{rule6}
\end{align}
\end{subequations}
where $A_i$ and $B_i$ are agents 
of group A and B, respectively, 
in state $S=i$.
For example, Eq. (\ref{rule3}) states 
that a member of the population $A_2$ can 
interact with a member of $A_3$ and that 
the result of the interaction are two 
members of the population $A_2$. 

The rules ($1a - 1d$) are introduced in order 
to emulate a process of imitation, where
interactions between agents of different groups 
lead to non preferential states of pronunciation, 
and interactions between agents of the same group,
conversely, reinforce the 
preferential states of the group.
In this respect, probabilities $q,p$ 
define the interaction strength
between agents of different groups, 
and the interaction strength
between agent of the same group, respectively.
Moreover, the noisy component expressed by rules $(1e)$ and $(1f)$,
captures the variations 
caused by both random phonetic changes
and the production of possible allophones
in both $PS^{A}$ and $PS^{B}$, respectively.
On the other hand, 
note that in the frame of the proposed 
imitation rules,
the changes in the states
occur only when the AP distance,
between the referent and the active agents 
is equal to one (i.e. $|\Delta S_{ij}|= 1$,
with $i,j$ reference and active agents).
The idea here is modeling the changes
in the context of close or similar sounds
\cite{pardo2006phonetic, ohala1990phonetics, nielsen1961implicit} 
between the different PSs, 
neglecting the contribution
to the changes of any other 
possible interactions.

In this theoretical frame, our proposal
has been inspired in the models
of opinion formation dynamics
\cite{castellano2009statistical, szabo2002three}
where social interactions,
in the context of a social debate,
drives the population to emergent states of
consensus or polarization
\cite{chacoma2015opinion, chacoma2018dynamical, chacoma2014critical}.

Macroscopically, 
in the frame of evolutionary dynamics, 
the global state of the system
can be analyzed by counting
the number of agents in $A,B$,
in the states $S=1,2,3$.
In the following section, based on this idea
we introduce the master equation of
the process and derive the 
evolution equations for
the first moments,
or mean-field approximation.

\section{The equations of evolution}
\label{sec:eq_ev}

Let $N^A_i$, $N^B_i$ be
the number of agents in $A,B$ in state $S=i$,
with $i=1,2,3$ (hereafter referred to as the occupation numbers).
The master equation of the system is given by,
\begin{equation}
\label{master}
\frac{\partial P(\vec{x};t)}{\partial t} = 
\sum_{\vec{y}\neq \vec{x}} T(\vec{x}|\vec{y})P(\vec{y};t)-
\sum_{\vec{x}\neq \vec{y}} T(\vec{y}|\vec{x})P(\vec{x};t),
\end{equation}
where $\vec{x}=(N^A_1, N^A_2, N^A_3 ; N^B_1, N^B_2, N^B_3 )$
is the so called occupation vector;
$P(\vec{x};t)$ is the probability to find the system 
with an occupation vector $\vec{x}$ at time $t$,
and $T(\vec{x}|\vec{y})$ is a transition probability 
from a global state given by an occupation vector $\vec{y}$ 
to another given by $\vec{x}$.
In this approach, we are considering 
a fixed population in both groups, 
therefore for all $t$ we have,

\begin{equation}
\label{Ns}
\begin{split}
N_A &= N^A_1+N^A_2+N^A_3, \\
N_B &= N^B_1+N^B_2+N^B_3,\\
N   &= N_A+N_B.
\end{split}
\end{equation}

From the master equation (\ref{master}), we can derive 
the evolutionary equations for the first moment
of the occupation numbers (mean-field approximation).
For instance, for $N^A_2$ we have,

\begin{equation}
\label{mean}
\frac{d}{dt} \avg{N^A_2} = 
\avg{T(N^A_2+1|N^A_2)} - 
\avg{T(N^A_2-1|N^A_2)},
\end{equation}
the transitions $T$ are defined by the rules proposed
in the last section, and depend on probabilities 
$p$, $q$, and $r$.
For the case by which the system
increases one agent in $N^A_2$,
$T$ can be written as

\begin{equation}
\label{eq:T1}
T(N^A_2+1|N^A_2) = 
\frac{N^A_1}{N} r +
\frac{N^A_3}{N}\frac{N^A_2}{N-1}p,
\end{equation}
where the first term is the probability
of finding an agent of group A in state $1$,
times the probability it randomly goes to state $2$;
and the second term is the probability of 
one interaction between an active agent of group $A$
in state $3$ and a reference agent of the 
same group in state $2$, 
leads to the former to imitate the 
latter with probability $p$.

For the case in which the occupation number $N^A_2$ 
decrease in one agent, the transition is given by,

\begin{equation}
\label{eq:T2}
T(N^A_2-1|N^A_2) = 
\frac{N^A_2}{N}\frac{N^B_3}{N-1}q + 
\frac{N^A_2}{N}r,
\end{equation}
where the first term shows the loss of an active agent 
of group A in state $2$ due to the interaction 
with a reference agent of group B in state $3$,
and the second term stands for the random loss of an
agent in $2$ who moves to $1$.

Replacing \ref{eq:T1} and \ref{eq:T2} in equation (\ref{mean}), 
we obtain,

\begin{equation}
\label{mean2}
\begin{split}
\frac{d}{dt} \avg{N^A_2} & = 
\avg{\frac{N^A_1}{N} r +
\frac{N^A_3}{N}\frac{N^A_2}{N-1}p} -
\\ &-
\avg{\frac{N^A_2}{N}\frac{N^B_3}{N-1}q + 
\frac{N^A_2}{N}r},
\\
\\
\frac{d}{dt} \avg{N^A_2} &= 
\frac{r}{N} \avg{N^A_1} +
\frac{p}{N(N-1)} \avg{N^A_3 N^A_2} -
\\ &-
\frac{q}{N(N-1)} \avg{N^A_2 N^B_3} - 
\frac{r}{N} \avg{N^A_2}.
\end{split}
\end{equation}

From now on, we will consider
the evolution of 
the numbers are uncorrelated, 
then $\avg{N^A_2 N^B_3}= \avg{N^A_2} \avg{N^B_3}$
and $\avg{N^A_3 N^A_2}= \avg{N^A_3} \avg{N^A_2}$.

Moreover, we re--scale the time
as $t \rightarrow \frac{t}{N}$, 
use the approximation for large
population $(N-1) \approx N$,
and define the occupation number as fractions 
of the total population:
$a_1 = \avg{\frac{N^A_1}{N}}$,
$a_2 = \avg{\frac{N^A_2}{N}}$,
$a_3 = \avg{\frac{N^A_3}{N}}$,
$b_3 = \avg{\frac{N^B_3}{N}}$.
Using the notation and approximations introduced above, 
we can write equation (\ref{mean2}) as follows,

\begin{equation}
\dot{a_2}=
r a_1 + p a_2 a_3 - q a_2 b_3 - r a_2
\end{equation}

Similarly, it is possible to obtain the 
equations for the evolution of  
the other occupation numbers,
which define the following system of coupled 
differential equations,

\begin{equation}
\label{ds}
\begin{split}
\dot{a_1} &= r ( a_2 - a_1)\\
\dot{a_2} &= - r ( a_2 - a_1) + p a_2 a_3 - q a_2 b_3\\
\dot{a_3} &= - p a_2 a_3 + q a_2 b_3\\
\dot{b_1} &= - p b_1 b_2 + qb_2 a_1\\
\dot{b_2} &=   p b_1 b_2 - q b_2 a_1 - r  (b_2 - b_3)\\
\dot{b_3} &= r  (b_2 - b_3)
\end{split}
\end{equation}
It is important to highlight
that since we are now working with
the fractions of the occupation 
numbers, the constraints related 
to the fixed population become:
$n_A =  a_1+a_2+a_3$, 
$n_B = b_1+b_2+b_3$, and
$n_A + n_B= 1$, 
where
$n_A = \frac{N_A}{N}$,
$n_B = \frac{N_B}{N}$.
The constraints also show that it is possible 
to reduce the rank of the system to four, 
however, for the sake of clarity, 
we decided to keep the six equations 
for a more detailed analysis.

Lastly, note that, 
if there are equilibrium points 
in system (\ref{ds}),
in the frame of our model,
these may be related to
the population reaching
consensus about a common 
general way of pronunciation
in both groups.
In the following section we will probe 
the existence of four equilibrium points, 
and will show that the stability of these
emergent states strongly depends 
on the interactions rates intra- and inter-group.

\begin{figure*}[t!]
\centering
\includegraphics[width=0.8\textwidth]{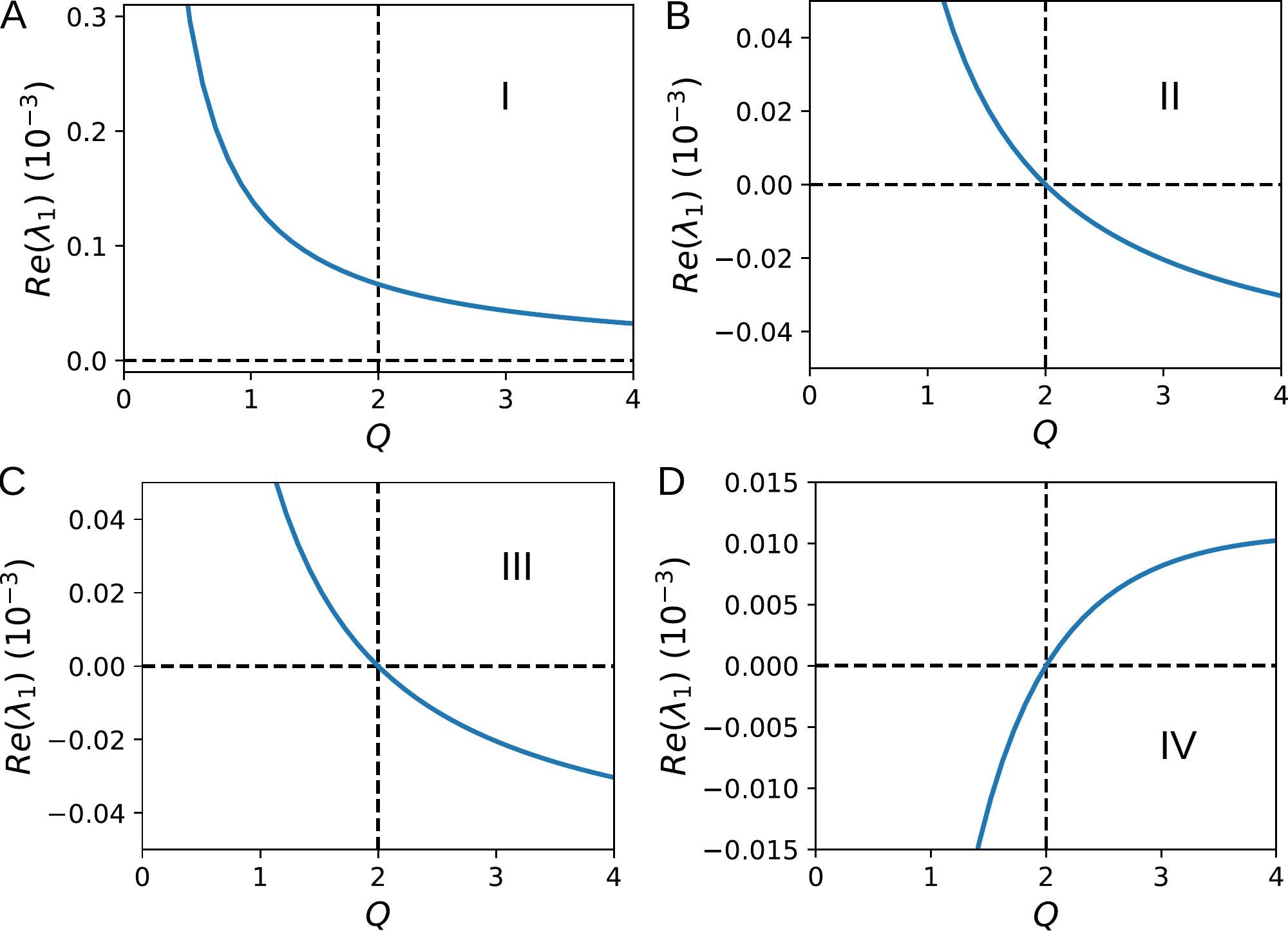}
\caption{Stability analysis. Real part of the 
largest non-trivial eigenvalue of the Jacobian Matrix ($\lambda_1$), 
as a function of the parameter $Q$.
Panels A,B,C and D show the numerical calculation
for the equilibria I, II, III and IV, respectively.
Note that when $Re(\lambda_1)>0$ the system is 
unstable, therefore equilibrium I will be unstable 
for all value of $Q$; II and III
will be unstable for $Q<2$, and IV for $Q>2$.
}
\label{autoval}
\end{figure*}

\begin{figure}[t!]
\centering
\includegraphics[width=0.43\textwidth]{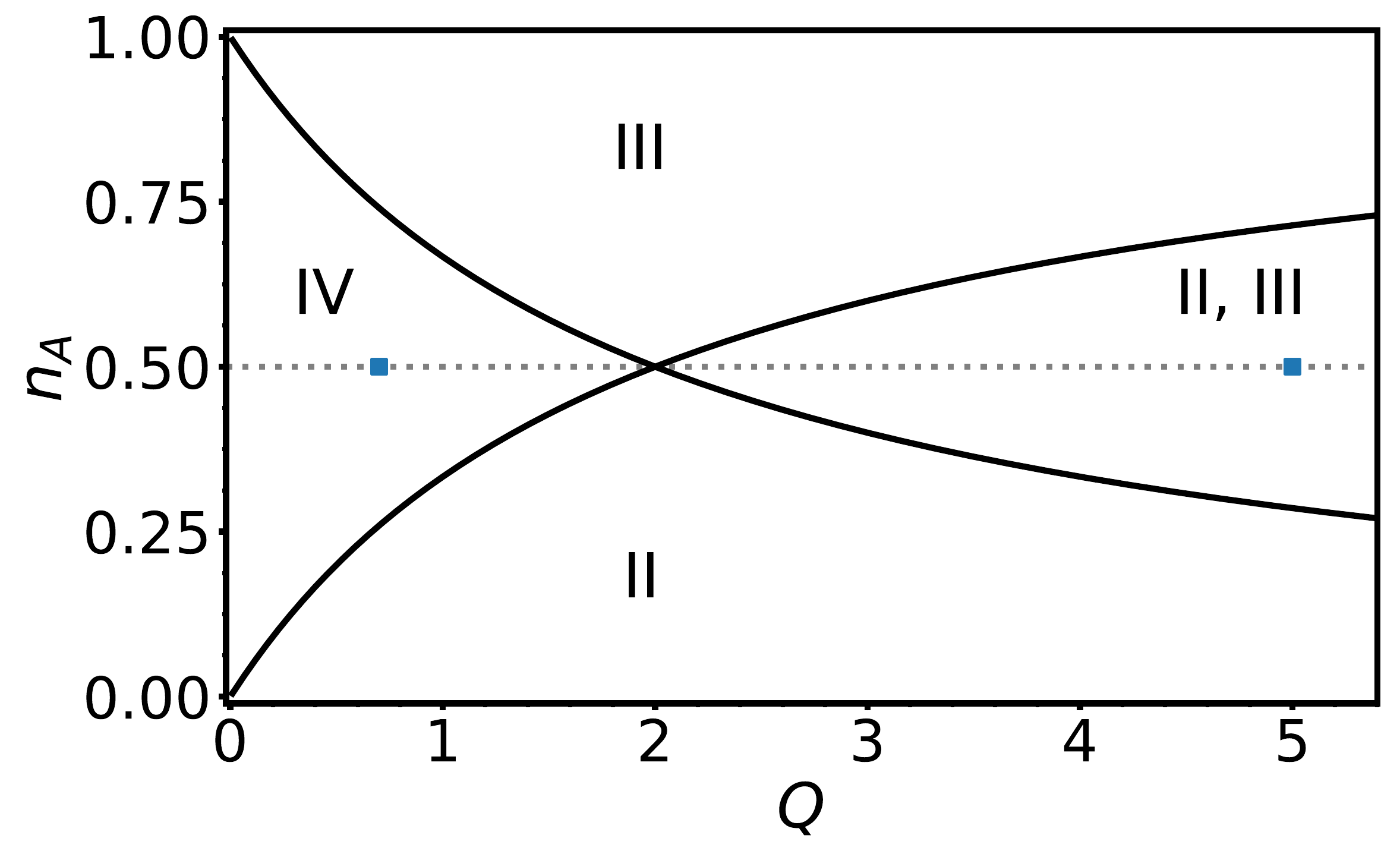}
\caption{Projection of the phase diagram in the plane $(n_A, Q)$. The curves $n_A = 2/(2+Q)$ and $n_A = Q/(2+Q)$ divide the plane into four regions with different stability. The labels for each region correspond to the equilibria that are stable inside the region. The two blue markers represent the parameters that were used in the main text.}
\label{fig:phase_diagram}
\end{figure}

\section{The emergent states}
\label{sec:estabilidad}

At time  $t\rightarrow \infty$, if an equilibrium exists,
it must satisfy $\dot{a_i}=0$ and $\dot{b_i}=0$.
In these conditions, from Eqs. (\ref{ds}) and
using the population constraints, it is possible to probe 
the existence of four non trivial equilibrium points,

\begin{enumerate}[I]
\item 
$\vec{n}_{eq} = (0,0, n_A; n_B, 0,0 )$
\item  
$\vec{n}_{eq} = (0,0, n_A; 0, \frac{n_B}{2}, \frac{n_B}{2} )$
\item      
$\vec{n}_{eq} = (\frac{n_A}{2},\frac{n_A}{2} ,0; n_B,0,0 )$
\item 
$\vec{n}_{eq} = (\alpha, \alpha, Q \beta; Q \alpha, \beta, \beta)$,
where 
$\alpha=\frac{2n_A-Qn_B}{4-Q^2}$, \\
$\beta =\frac{2n_B-Qn_A}{4-Q^2}$ and $Q = \frac{q}{p}$.
\end{enumerate}

The first equilibrium can only be reached if 
the initial conditions are set at this point,
later we will show this equilibrium is unstable.
Equilibria II and III are the cases
where the system loses the back and the 
frontal AP, respectively.
Equilibrium IV, on the other hand,
shows a mixed final state, where
the system reaches a balance
among the three states of pronunciation.

The stability analysis 
around the equilibrium can
be performed by analysing the 
eigenvalues of system (\ref{ds})
(see Appendix \ref{apendice}).

\begin{figure*}[t]
\centering
\includegraphics[width=0.8\textwidth]{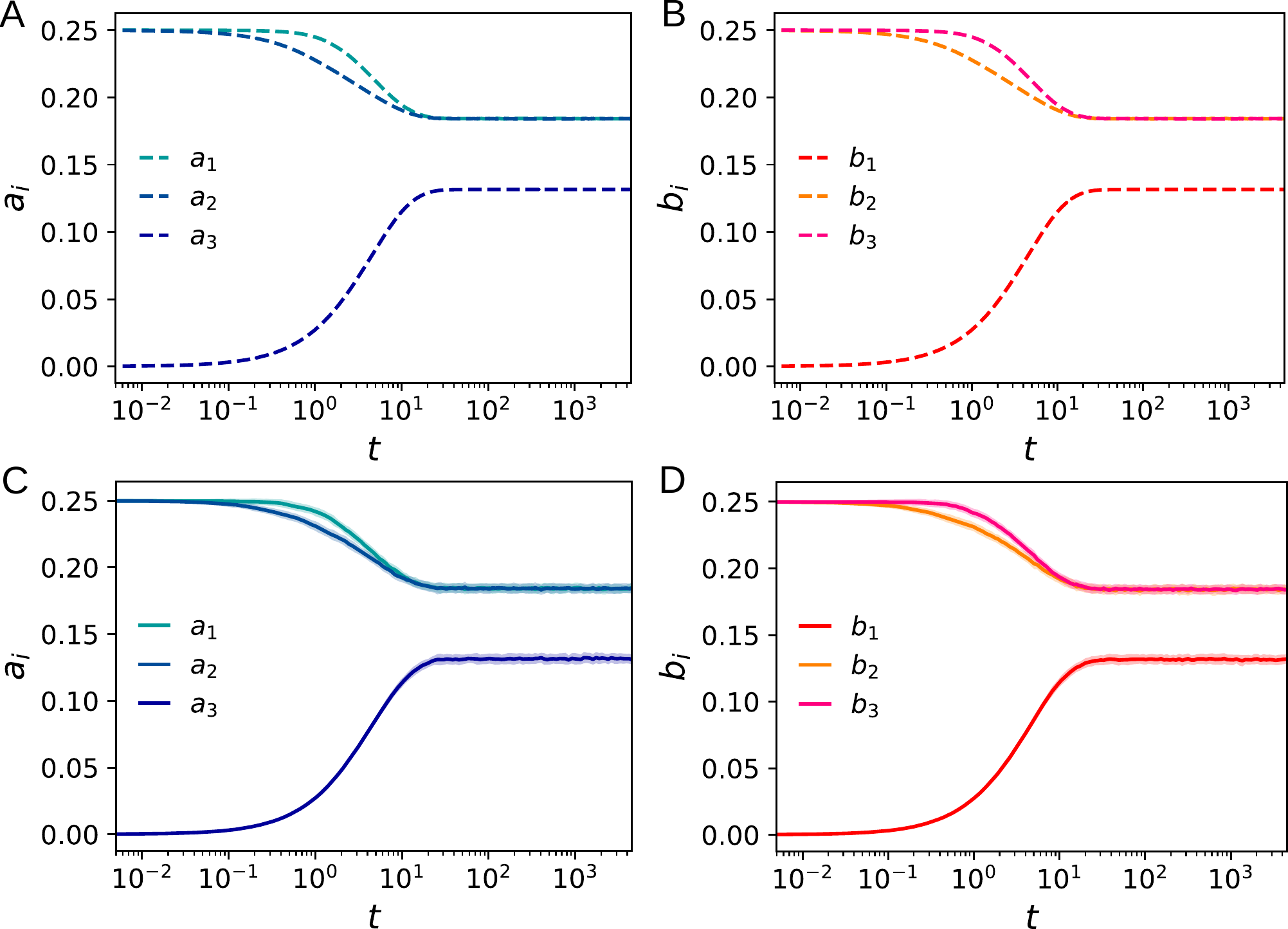}
\caption{
The case of equilibrium IV stable.
For $Q=0.7$, panels A and B show the first moment evolution 
of the numbers related
to group A ($a_1,a_2,a_3$) and B 
($b_1,b_2,b_3$), respectively.
The curves are the outcome 
of the evolution  of Eqs. (\ref{ds}),  
which were solved by performing 
a Runge Kutta 8th order method.
Panels C and D show the
average curves over $100$ realizations,
of the numbers in the ABM. The shadow indicate 
the standard deviation around the mean value.
In these conditions, the population evolves to equilibrium IV,
a mixed state of pronunciation where 
the system reaches 
$\vec{n}_{eq} 
= (\frac{5}{27}, \frac{5}{27}, \frac{7}{54}; \frac{7}{54},\frac{5}{27},\frac{5}{27}) \approx (0.18, 0.18, 0.13; 0.13, 0.18, 0.18)$.
}
\label{Q07}
\end{figure*}

We are particularly interested in studying
the evolution of the system as a function of $Q$
since this parameter controls the interaction 
between $A$ and $B$, and also
rules the population state of the equilibrium point IV.
To this purpose, we
set a constant equal population $n_A=n_B$, 
$r=0.5$, $q=0.5$ and vary the parameter $p$,
in order to focus only in the 
study of the effect of $Q = q/p$, on the equilibrium 
conditions of the system.
In this frame, we have calculated
the eigenvalues as a function of $Q$,
and studied the sign of the real part of the
eigenvalues to evaluate the stability conditions.

The plots in figure \ref{autoval} show 
the curves for the real part of the largest non--trivial
eigenvalue, $Re(\lambda_1)$ vs. $Q$ for the 
four equilibrium points.
Panel A in this figure shows the calculation for state I.
We can see that for all $Q$ there is an
eigenvalue with positive real part, which means
this  equilibrium state is always unstable.
In panels B and C we can observe that states II and III behave similarly each other.
This is expected
because these two states are symmetric.
States II and III are unstable for $Q<2$ 
since they have at least one eigenvalue with
positive real part. Finally, panel D shows the calculation for
equilibrium state IV; in this case,
complementary to states II and III,
the equilibrium is unstable
for $Q>2$.

Clearly, coefficient $Q$, which measure 
the relative intra- and inter-groups imitation rates,
determines the stability of the different 
equilibrium states of the system.
We can understand this phenomenon by 
reasoning as follows: 
At $Q=2$, the inter-group interactions 
equals the sum of the interaction rates 
within each group 
($Q=2 \rightarrow q= p+ p$),
and equilibrium IV becomes
$\vec{n}_{eq} = 
(\alpha, \alpha, Q \beta; Q \alpha, \beta, \beta)=
(\frac{1}{8}, \frac{1}{8}, \frac{1}{4};
\frac{1}{4},\frac{1}{8},\frac{1}{8})$,
i.e. there is a balance between the number 
of agents in preferential and non-preferential 
states of pronunciation in both groups
($n^A_1+n^A_2=n^A_3$ and $n^B_1=n^B_2+n^B_3$).
For $Q>2$, the system
loses the balance, 
the number of agents in non-- preferential
states becomes larger than the number of agents in 
preferential states, hence equilibrium IV 
becomes unstable and depending on the initial conditions the system evolves to equilibrium II or III. 
Furthermore, in the stochastic version of the model 
and balanced initial conditions fluctuations can 
determine the final equilibrium state, 
which occurs with equal probability for states II and III. 

Finally, a more general stability analysis of 
the mean field model, which
includes unbalanced conditions $n_A \neq n_B$,
is conducted in  Appendix \ref{apendice}. The main results of this
analysis is summarized in the phase diagram shown in Fig.~\ref{fig:phase_diagram}. There are
four stability regions in the $n_A$ - $Q$ plane. One for point IV at low values of Q. Another two
for points II and III, respectively, and one region where point II and III coexist. All
our analysis will be restricted to two representative values of
$Q$ in this phase diagram.

\begin{figure*}[!t]
\centering
\includegraphics[width=0.8\textwidth]{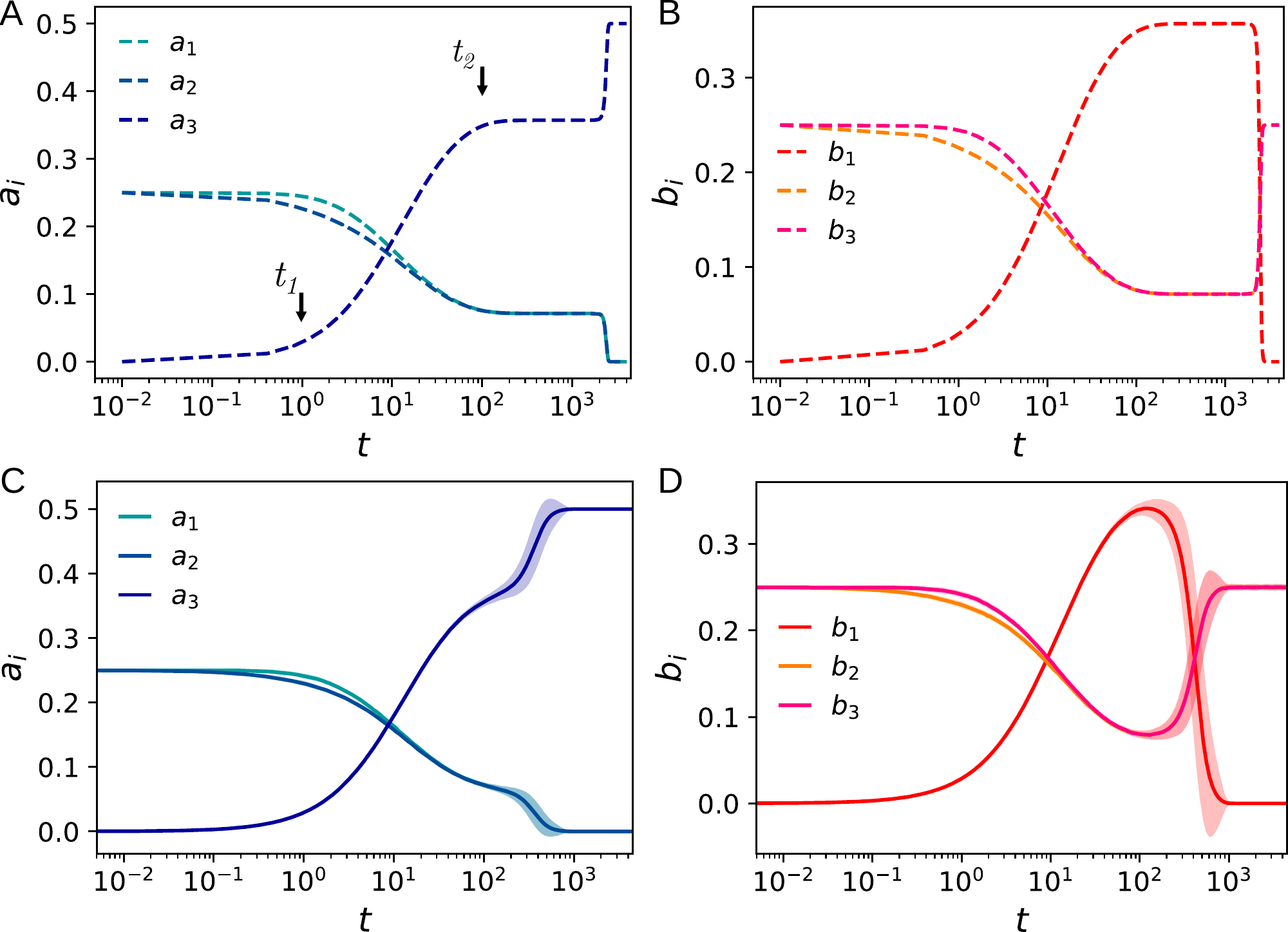}
\caption{
The case of equilibrium II stable.
For $Q=5$, panels A and B show the first moment evolution 
of the numbers related to group A and B, respectively.
Panels C and D, on the other hand, 
show the average curves over 100 realizations,
of the numbers in the ABM. The shadow in the plots
indicate the standard deviation whch we see it increases 
when the system reach the unstable equilibrium IV.
Note how at the beginning the system seems
to evolve toward IV,
but escape toward the equilibrium II, which is
stable in these conditions 
($\vec{n_{eq}} = 
(0,0,\frac{1}{2}; 0,\frac{1}{4},\frac{1}{4})$).
The arrows in panel A, indicate
the time scales obtained from the eigenvalues 
analysis: $t_1 \approx 1$ and $t_2 \approx 10^2$.
}
\label{Q5}
\end{figure*}

In order to visualize the time evolution 
of the occupation number in the system,
and to study the convergence to the equilibrium points,
we solved numerically the set of coupled 
differential equations given in 
Eqs. (\ref{ds}),
and also performed numerical simulations of 
the stochastic agent based model (ABM)---using the rules proposed in section 
\ref{sec:model}---in order to test the effect
of fluctuations in the dynamics of the system.
In a first approach 
we explore two limit cases far for the 
singular point $Q=2$: 
the case of $Q=0.7$ where the inter--group
interaction domain the dynamics,
and the case of $Q=5$ where, conversely,
the dynamic is governed by the intra--group 
interactions.
For this purpose, we set
(i) the initial conditions on
$\vec{n}(t=0) = 
( \frac{n_A}{2}, \frac{n_A}{2}, 0; 
0,\frac{n_B}{2}, \frac{n_B}{2} )$
(preferential states for both groups),
(ii) the fraction  
of agents such that $n_A=n_B$ 
(equal population in both group),
(iii) the parameters $r=0.5$ and $q=0.5$,
and (iv) for the ABM simulations, a 
population size of $N= 10^4$.
The results are
summarized in the plots of figure \ref{Q07} and \ref{Q5},
which we describe in the following.
Figure \ref{Q07} shows 
the evolution of the occupation numbers
for $Q=0.7$ where
state IV is stable.
Panels A and B show the evolution
in the mean field approximation (Eqs. (\ref{ds})),
whereas panels C and D show the outcome 
of the ABM; here we have 
averaged the results over
$100$ performed simulations.
We can see that, as expected,
the system evolves toward equilibrium IV
and the mean reaches the value
$\vec{n}_{eq} 
= (\alpha, \alpha, Q \beta; Q \alpha, \beta, \beta)
= (\frac{5}{27}, \frac{5}{27}, \frac{7}{54}; \frac{7}{54},\frac{5}{27},\frac{5}{27})$.

Figure \ref{Q5},
by contrast, 
shows the evolution for $Q=5$ where 
equilibria II and III are stable;
in this case, the realization 
showed in the plots went to equilibrium II.
Notably, at the beginning of the evolution,
the system seems to stabilize
at equilibrium IV, 
but at larger times moves toward equilibrium II, as expected.
The arrows in figure \ref{Q5} panel A,
indicate the relaxation times operating
at each regime, given by the 
inverse of the eigenvalues ($\propto 1/\lambda_i)$.

\begin{figure*}[t!]
\centering
\includegraphics[width=0.8\textwidth]{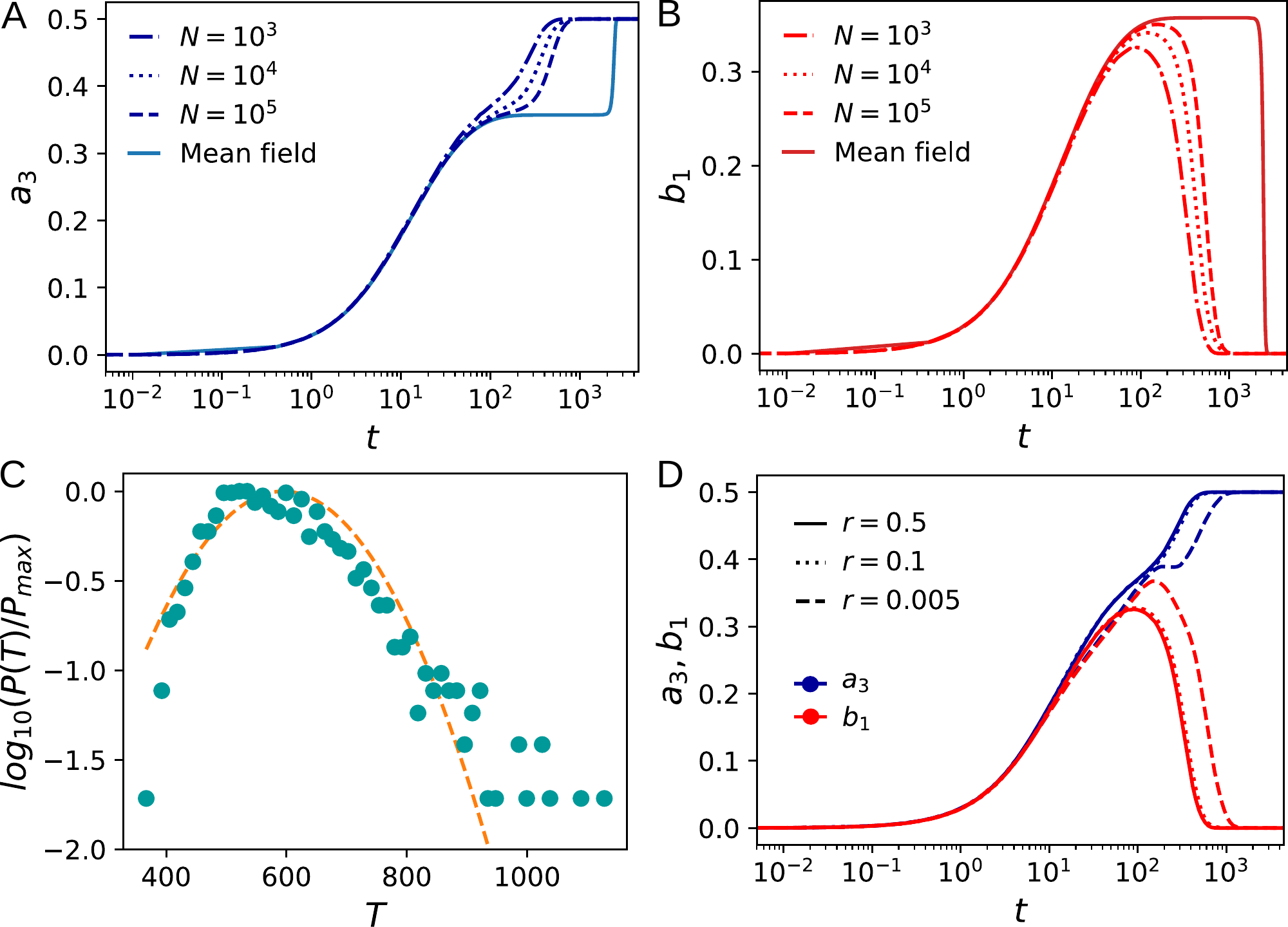}
\caption{
Panel A and B, the average curves for the evolution of the 
numbers $a_3$ and $b_1$, respectively, for different
values of the population size $N$, compared with the 
mean field approximation in order to see the finite size effect.
Panel C, distribution of final times $T$, needed to
the system reach equilibrium II at $Q=5$.
The dashed line in the background indicate 
a Gaussian distribution with the same 
mean and standard deviation, which helps to visualize
the presence of a large tail in distribution $P(T)$.
Panel D, the average curves for the evolution of the 
numbers $a_3$ and $b_1$ for different values of the
parameter $r$.
}
\label{dist_T}
\end{figure*}


In a second approach, we explore 
finite size effects in the system,
by performing ABM simulation
for different sizes ($N$) of the total population 
and comparing with the mean field approximation.
For the case $Q=5$, we can see from 
figure \ref{dist_T} panels A and B,
that the finite size effects become notable
once the system reaches
the unstable equilibrium IV.
The larger the population 
the lower the fluctuations, therefore
the system will spend more time at 
the unstable equilibrium IV
before fluctuations lead it to the stable state II.

In connection with the expressed above,
we measured the distribution
of times ($T$) needed for the system 
to reach the final equilibrium at $Q=5$  
by performing $2000$ realizations 
of the stochastic model.
Figure \ref{dist_T} panel C,
shows the distribution obtained.
We can see a non symmetric distribution with 
a peak around $T \approx 600$,
and a tail at the right of the distribution.
As we said before, in every realization
the system stays in the unstable 
equilibrium IV before it reaches the final stable state, 
therefore the total time $T$ depends on the magnitude
of the fluctuations which drive the system 
from the unstable equilibrium 
to the stable one. The latter explains the
tail at the right of the distribution and
also the differences in the final times observed 
among the stochastic simulations and the mean field approach.

On the other hand, as we show above the equilibrium 
points are not dependent of parameter $r$;
however, since $r$ is a stochastic 
parameter of the model,
the dynamics towards 
the equilibrium should depend on it.
In Panel D on figure \ref{dist_T} 
we show the evolution of the numbers
$a_3$ and $b_1$ for $r = 0.5,\ 0.1, \ 0.005$.
For the lowest value explored ($r=0.005$),
we can see a different behaviour 
around the unstable equilibrium IV,
which seems to indicate a link
between the reduction of the stable
attractor force and the reduction of $r$. 

\begin{figure*}[t!]
\centering
\includegraphics[width=0.8\textwidth]{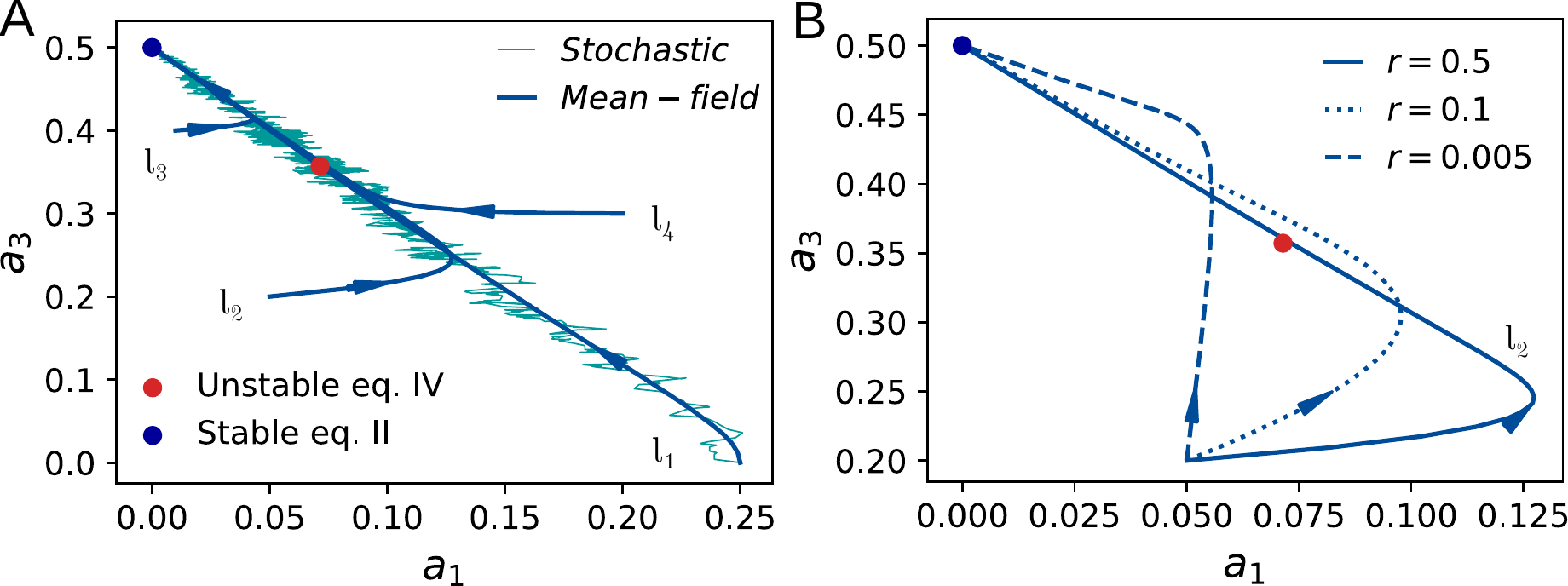}
\caption{
Trajectories in the plane $a_1 - a_3$.
Panel A, changing the initial conditions, keeping $r=0.5$.
Panel B, changing parameter $r$, for trajectory 
starting from the same initial condition than $l_2$
in panel A. The blue lines correspond to the 
solution of the differential equations, 
while the cyan curve in Panel A corresponds 
to a single realization of the stochastic 
simulation, starting from the initial condition of 
trajectory $l_2$.
}
\label{a1_a3_2}
\end{figure*}

Lastly, in order to complete a global analysis, 
we test the system behaviour under 
changes in the initial conditions.
Figure \ref{a1_a3_2} panel A shows 
trajectories in the plane $a_1 - a_3$ for $Q=5$,
obtained from the mean field approach simulations, 
where we have tried different initial conditions
for the numbers $a_i$, keeping $r=0.5$
(the different trajectories in the plane
are indicating in the plot as $l_i$, $i=1,2,3,4$).
We can see that depending on 
the proximity to the stable attractor II
the system will explore the unstable equilibrium IV,
as in the case of trajectories 
$l_1$, $l_2$ and $l_4$, or it will not as in 
the case of trajectory $l_3$. 
For the case of $l_1$ we have additionally plotted 
the trajectory obtained from a single stochastic realization.

The plot in figure \ref{a1_a3_2} panel B, 
on the other hand, complements the information given 
in figure \ref{dist_T} panel D, showing 
the effect of parameter $r$ on the 
trajectories in the plane $a_1 - a_3$.
In this regards,
the parameter $r$ regulates the velocity
of the transitions $\ce{A_1 <=> A_2}$ 
(and also $\ce{B_2 <=> B_3}$). 
When $r$ is strong enough, these transitions 
occur much faster than the rest and thus,
populations $A_1$ and $A_2$ tend 
to have the same number of individuals, 
independently on the initial conditions. 
This aspect explains why all the trajectories 
in Figure \ref{a1_a3_2} panel A rapidly 
approximate to the line $a_3 = -2a_1$. 
If the system starts from an initial 
condition with a low value of $a_3$, 
then the trajectory will be forced
to go through equilibrium IV, 
as it can be seen in the figure. 
When we reduce the value of $r$,
it is harder for the system to equal the values
of $a_1$ and $a_2$ and thus, the trajectories 
in general deviate from the line and 
equilibrium IV is avoided,
as we show in Figure \ref{a1_a3_2} panel B.

In the frame of the proposed model,
state II can be related to the change 
/\textphi/$\rightarrow$/h/ in Castilian,
and state III to the observed in other
Romance languages like Portuguese or Catalan.
Therefore, for $Q>2$, the model seems
to capture very well 
the current pronunciations
that emerged---from the real LS---in the Iberian peninsula.
On the other hand, equilibrium IV, 
describes an emergent state of mixed 
pronunciation, which means there 
are agents using different AP
to pronounce the same word.
This is rarely observed in 
the real case, but can be used
to understand the existence of some 
regionalism or local accents in the peninsula
\cite{martinez1997issues}.

\section{Conclusions}

In this work, we have proposed
a model to study phonetic
changes in the AP used for pronounce a single word 
as an evolutionary process 
guided by social interaction of imitation
between two groups of people
with different phonological systems.
Inspired by the case of the change /\textphi/$\rightarrow$/h/ 
in Castilian language, 
we have studied a fixed population
made up of two groups of interacting people,
A and B, such that
group A has a trend to produce 
frontal fricatives and conversely
group B has a trend to produce back 
fricatives.
The rules of the model were proposed 
based on empirical observations,
and were thought to link the
phonetic changes with a process 
of social interactions inter- and intra-group.
The model was mathematically formalized
in section \ref{sec:model}
as a stochastic process
where the variable $S\in\{1,2,3\}$, representing the AP
for every agent in the population,
changes according to the proposed interaction rules.

In this frame we studied
the temporal evolution of the occupation numbers,
and from first principles,
we derived the coupled system of differential equations
which defines the dynamics in the mean field approximation.
In the equilibrium,
we found three non trivial final states,
which we have related to 
the emergence of general states of consensus
in the way a word is pronounced.
In this regards, we found that when 
the rate of interaction among 
agents from different groups 
becomes larger than
the sum of the rates within each group
($q > 2 p$),
the system exhibits two emergent
states (equilibria II and III) which 
capture very well both, the 
middle-back pronunciation used in Castilian
and the front-middle pronunciation 
observed in other Romance languages,
as the case of Portuguese.
From a social point of view,
we can link the condition
$q > 2 p$ 
to the situation where the relation 
among individuals from
different groups is larger enough
to allow a common general consensus
in spite of the cultural differences.

The model we have introduced is based on a mean field approach since all agents
interact among them, hence it does not consider the influence
of the structure of social interactions in reaching a consensus
for the phonetic changes. This is of course a simplification, as it is
known that the structure of social interactions strongly influences human
behavior and the evolution of social and cognitive processes~\cite{baronchelli2013networks}. Even more, human mental lexicon is supposes to be assembled according to a multiplex 
network structure~\cite{stella2016mental}. Hence, one can expect the network structure to play a key role in any particular dynamics of phonetic changes.
However, Baxter, et al.~\cite{baxter_blythe_croft_mckane_2009} proved that in several neutral interactor models inspired
in Trudgill's theory for the emergence of New Zealand English,
the structure of the underliying social network has a minor effect
in the final state distribution of speaker's grammar (linguemes) produced
by these models. According to Trughill's deterministic theory, frequency of use
and accomodation are the only factors to be taken into account for the prevalence
of a given lingueme. Our model is deterministic in terms of its parameters.
The final phonetic state is correlated to the initial frequency of agents
using a given phonetic system for the case $Q>2$ (states II and III), in those states
one phonetic system prevails over the other.
However, the proposed imitation mechanism that rules the  interaction between agents
of different phonetic systems, can lead to a final state in which the two phonetic
systems coexist. This is particularly relevant in the analysis of Castilian phonetic
when the system is considered as a whole. In this case the LS contains fricative
words that keep the Latin pronunciation as well as other words that change to
glottalization, which means that there is a coexistence of the articulation
places in the global system.

Finally, more realistic models must consider the structures of social interactions
in the dynamics. Then, it would be interesting to analyze the interplay between
the network topology and the dynamics of phonetic changes generated by our model.
In this regards, we let as an open problem to be faced in future works the
study of the effect of the structure of social interactions in phonetic changes.

\appendix

\section{Stability analysis}
\label{apendice}

In this appendix we perform a
stability analysis to extend the discussed 
in section \ref{sec:estabilidad}.

Equation system (\ref{ds}) can be reduced to a four-dimensional system using the constraints $a_1 + a_2 + a_3 = n_A$ and $b_1 + b_2 + b_3 = n_B$. Defining the parameters $\omega = r / p$, $Q = q/p$ and scaling the time as $\tau = pt$, we have

\begin{equation}
\begin{split}
\dot{a_1} &= \omega (n_A - 2 a_1 - a_3)\\
\dot{a_3} &= (Q b_3 - a_3) (n_A - a_1 - a_3)\\
\dot{b_1} &= (Q a_1 - b_1) (n_B - b_3 - b_1)\\
\dot{b_3} &= \omega (n_B - 2 b_3 - b_1)
\end{split}
\end{equation}

where we have changed the notation $\dot{x} \equiv dx/d\tau = (1/p) dx/dt$.

The Jacobian of the reduced system is

\begin{widetext}
  \begin{equation} \label{eq:Jacobian}
    J = 
    \begin{pmatrix}
    -2 \omega & -\omega  & 0 & 0 \\
    a_3 - b_3 Q & a_1 + 2 a_3 - n_A - b_3 Q & 0 & (n_A - a_1 - a_3) Q \\ 
    (n_B - b_3 - b_1) Q & 0 & b_3 + 2 b_1 -n_B -a_1 Q & b_1 - a_1 Q \\
    0 & 0 & -\omega & -2 \omega
    \end{pmatrix}.
  \end{equation}
\end{widetext}

We proceed now to analyze the stability of each equilibrium. Before we start, it is important to notice that, as we have re-scaled the time as $\tau = p t$, the eigenvalues of the original system can be expressed as $\lambda = p \mu$, where $\mu$ are the eigenvales of the reduced system.

Evaluating \ref{eq:Jacobian} at equilibrium I and computing its eigenvalues, we have

\begin{align}
    \mu^B_{\pm} &= \dfrac{n_B -2\omega \pm \sqrt{(n_B-2\omega)^2+4n_B\omega}}{2}, \\
    \mu^A_{\pm} &= \dfrac{n_A -2\omega \pm \sqrt{(n_A-2\omega)^2+4n_A\omega}}{2}. \nonumber
\end{align}

The four eigenvalues are always real, and two of them (the ones with the plus sign) are always positive. This means that equilibrium I is always unstable, and thus, it lacks of physical interest.

For equilibrium II, the corresponding eigenvalues are

\begin{align}
    \mu_{\pm} &= \dfrac{2 n_A -Q n_B -4 \omega \pm \sqrt{(2n_A-Q n_B )^2 + 16 \omega^2}}{4}, \nonumber \\
    \mu_3 &= - \dfrac{n_B}{2}, \\
    \mu_4 &= - 2\omega. \nonumber 
\end{align}

The eigenvalues $\mu_-$, $\mu_3$ and $\mu_4$ are always negative, but $\mu_+$ can be either positive or negative. The region of the parameter space where it is negative (and thus, where the equilibrium is stable) is given by

\begin{equation}
    Q > \dfrac{2 n_A}{n_B}
\end{equation}

Considering $n_B = 1 - n_A$, the previous inequality can be solved for $n_A$ as 

\begin{equation} \label{eq:EqII_stability}
    n_A < \dfrac{Q}{2+Q}.
\end{equation}

Equilibrium III is symmetric with respect to equilibrium II and its eigenvalues are

\begin{align}
    \mu_{\pm} &= \dfrac{2 n_B -Q n_A -4 \omega \pm \sqrt{(2n_B-Q n_A )^2 + 16 \omega^2}}{4},\nonumber \\
    \mu_3 &= - \dfrac{n_A}{2}, \\
    \mu_4 &= - 2\omega. \nonumber
\end{align}

Thus, the equilibrium is stable when

\begin{equation}
    Q > \dfrac{2 n_B}{n_A},
\end{equation}

which can be also expressed as 

\begin{equation} \label{eq:EqIII_stability}
    n_A > \dfrac{2}{2+Q}.
\end{equation}

Equilibrium IV exists only if $\alpha$ and $\beta$ are simultaneously greater than zero. It can be shown that the condition for this to happen is

\begin{align} \label{eq:EqIV_existence}
    \left(n_A - \dfrac{2}{2+Q}\right) \left(n_A - \dfrac{Q}{2+Q} \right)< 0.
\end{align}

To analyze the stability, lets first consider the particular case $n_A=n_B$. In this case, the corresponding eigenvalues are 

\begin{align}
   \mu_{1,\pm} =& \dfrac{-\left[n_A + 2\omega(2 + Q)\right]}{2(2+Q)} \nonumber \\
   &\pm \dfrac{\sqrt{\left[n_A + 2\omega(2 + Q)\right]^2 - 4\omega n_A (4-Q^2)}}{2(2+Q)}, \\
   \mu_{2,\pm} &= \dfrac{-\left[n_A + 2\omega(2 + Q)\right]}{2(2+Q)} \nonumber \\
   &\pm \dfrac{\sqrt{\left[n_A + 2\omega(2 + Q)\right]^2 - 4\omega n_A (2+Q)^2}}{2(2+Q)}. \nonumber
\end{align}

From these eigenvalues, $\mu_{1,\pm}$ and $\mu_{1,-}$ are always negative while $\mu_{1,+}$ is negative if and only if $Q<2$. For the general case $n_A \neq n_B$, we could not get explicit expressions for the stability analysis of this equilibrium. Instead, we performed simulations starting from different combinations of parameters and we found that, as long as condition \ref{eq:EqIV_existence} is satisfied, equilibrium IV is stable for $Q<2$ and unstable for $Q>2$.

Using Eqs \ref{eq:EqII_stability}, \ref{eq:EqIII_stability} and \ref{eq:EqIV_existence}, and taking into account the discussion in the previous paragraph, we can draw a phase diagram for our model in the plane $(n_A, Q)$, as we show in Fig.~\ref{fig:phase_diagram}.


\begin{acknowledgments}
This work was partially supported by grants from CONICET (PIP 112 20150 10028), FonCyT (PICT-2017-0973), SeCyT–UNC (Argentina) and MinCyT C\'ordoba (PID PGC 2018). 

\end{acknowledgments}

\end{document}